\documentclass[twocolumn]{aastex6}

\newcommand{\psrb}{PSR\,B1259-63/LS\,2883}
\newcommand{\pulsar}{PSR\,B1259-63}
\newcommand{\cstar}{LS\,2883}
\newcommand{\gammaray}{$\gamma$-ray}

\newcommand{\gaga}{$\gamma\gamma$}
\newcommand{\hess}{H.E.S.S.}
\newcommand{\fermi}{\emph{Fermi}-LAT}

\begin{document}


\title{Gamma-gamma absorption in the $\gamma$-ray binary system \psrb}

%% Use \author, \affil, plus the \and command to format author and affiliation 
%% information.  If done correctly the peer review system will be able to
%% automatically put the author and affiliation information from the manuscript
%% and save the corresponding author the trouble of entering it by hand.
%%
%% The \affil should be used to document primary affiliations and the
%% \altaffil should be used for secondary affiliations, titles, or email.

%% Authors with the same affiliation can be grouped in a single
\author{Iurii Sushch\altaffilmark{1}}
\affil{Centre for Space Research, North-West University, 2520 Potcheftroom, South Africa \\
DESY, D-15738 Zeuthen, Germany\\
Astronomical Observatory of Ivan Franko National University of L'viv, vul. Kyryla i Methodia, 8, 79005 L'viv, Ukraine}

\and

\author{Brian van Soelen\altaffilmark{2}}
\affil{Department of Physics, University of the Free State, 9300 Bloemfontein, South Africa}


\altaffiltext{1}{{iurii.sushch@desy.de}}
\altaffiltext{2}{vansoelenb@ufs.ac.za}

\begin{abstract}
{The observed TeV light curve from the $\gamma$-ray binary \psrb\ shows a decrease in the flux at periastron which has not been fully explained by emission mechanisms alone. This observed decrease can, however, be explained by \gaga\ absorption due to the stellar and disk photons. We calculate the \gaga\ absorption in \psrb\ taking into account photons from both the circumstellar disk and star, assuming the $\gamma$ rays originate at the position of the pulsar. The \gaga\ absorption due to the circumstellar disk photons produces a $\approx14\%$ decrease in the flux, and there is a total decrease of $\approx52\%$ ($>1$\,TeV) within a few days before periastron, accompanied by a hardening of the $\gamma$-ray photon index. While the \gaga\ absorption alone is not sufficient to explain the full complexity of the H.E.S.S. $\gamma$-ray light curve  it results in a significant decrease in the predicted flux, which is co-incident with the observed decrease. In addition, we have calculated an upper-limit on the \gaga\ absorption, assuming that the emission is produced at the apex of the bow shock. Future observations with CTA during the 2021 periastron passage may be able to confine the location of the emission based on the degree of \gaga\ absorption as well as measure the hardening of the spectrum around periastron.}
\end{abstract}


\section{Introduction} \label{sec:intro}

The binary system \psrb\ is a member of a small but growing class of sources known as \gammaray\ binaries which at the moment comprises of only six known objects.
These systems consists of a compact object, believed to be either a pulsar or  black hole, in orbit around a early type O or B star, and show a spectral energy distribution which peaks above $1$\,MeV \citep[see][for a detailed review of these systems]{2013A&ARv..21...64D}. The most recently discovered system lies in the Large Magellanic Cloud (LMC), the first to be detected outside of the Milky Way \citep{2016ApJ...829..105C}. This source has been detected by {\it Fermi}-LAT but, unlike the other sources, has not been detected at TeV energies. One other candidate \gammaray\ binary system, PSR\,J2032$+$4127/MT91\,213, which contains a $\gamma$-ray pulsar, has also recently been highlighted by \cite{2017MNRAS.464.1211H}. A possible TeV counterpart of this binary system, TeV\,J2032$+$4130, was one of the first sources ever detected at TeV energies and the first TeV source with no obvious counterpart at other wavelengths \citep{2002A&A...393L..37A}. The source, however, exhibits a steady TeV flux with no evidence of periodic behaviour and was at first suggested to be the  evolved pulsar wind nebula of PSR\,J2032$+$4127  \citep[][and references therein]{2014ApJ...783...16A}. The recent discovery that the pulsar is in a $\sim30$~year orbit around MT91\,213, and is expected to pass through periastron by the end of 2017, challenges the current interpretation \citep{2015MNRAS.451..581L, 2017MNRAS.464.1211H}.

Unlike the other $\gamma$-ray binaries, where the nature of the compact object is unknown, in \psrb\ it is known to be a 48\,ms pulsar, which is in an eccentric $\sim3.4$ year orbit around the companion \citep{1992MNRAS.255..401J, 1992ApJ...387L..37J,2014MNRAS.437.3255S}.  For the other systems both pulsar and black hole scenarios have been widely discussed. The companion, \cstar, is a Be star which is surrounded by a circumstellar disk, which is a region of enhanced stellar outflow \citep{1992MNRAS.255..401J, 1992ApJ...387L..37J,2011ApJ...732L..11N}. Circumstellar disks around Be stars are known to generate infrared (IR) emission produced mainly through free-free radiation,
which provides an additional target photon field for inverse Compton (IC) scattering \citep{2012MNRAS.426.3135V,2012ApJ...752L..17K} and \gaga\ absorption \citep{2014JHEAp...3...18S,2007Ap&SS.309..333O}.

The disk is thought to be inclined with respect to the binary orbit, and the pulsar crosses it twice each orbit.
The dense medium of the disk plays an important role in the resulting emission from the system. 
When the pulsar goes behind the disk its pulsed radio emission disappears, at about $t_{\mathrm{p}} - 20$\,d, and it reappears again after the second crossing of the disk, at about $t_{\mathrm{p}} + 15$\,d, where $t_{\mathrm{p}}$ is the time of periastron. In this way the location of the disk is determined.

As the pulsar approaches periastron, there is an increase in the unpulsed non-thermal radio, X-ray and TeV emission.
The radio, X-ray and TeV light curves show a similar behaviour, exhibiting a double-peak structure (or a hint of a double-peak structure as in case of the TeV light curve), with two asymmetrical peaks roughly coinciding with the time that the pulsar passes through the disk \citep[see][and references therein]{2014MNRAS.439..432C}. 
This suggests a possible connection between the equatorial disk and the variability of the non-thermal emission. The observed shape of the GeV light curve, however, strongly
deviates from those at other wavebands, with only faint (if any) emission detected at periastron and a sudden and very powerful flare starting from approximately 30 days after periastron \citep{2011ApJ...736L..11A, 2011ApJ...736L..10T, 2015MNRAS.454.1358C, 2015ApJ...798L..26T, 2015ApJ...811...68C}. This flare does not occur at the same time as the post-periastron peaks detected at other wavelengths \citep[e.g.][]{2015MNRAS.454.1358C}.
The flare is now known to be periodic as it has been detected around two consecutive periastron passages, though its cause is still not clear.

At TeV energies \psrb\ has been observed by \hess\ around four periastron passages, namely, 2004 \citep{2005A&A...442....1A}, 2007 
\citep{2009A&A...507..389A}, 2010 \citep{2013A&A...551A..94H}, and 2014 \citep{2015arXiv150903090R}. 
The TeV emission shows a clear variable behavior around the periastron passage, with a suggestion of a double peak structure, and a local minimum in the flux at periastron.
The double-peak shape of the TeV light curve would suggest a hadronic scenario
for the generation of the TeV emission. Indeed, if the pulsar wind is predominantly proton-loaded, one would expect two sharp peaks 
when the protons interact with the dense disk environment \citep{2007Ap&SS.309..253N}. The hadronic scenario could also explain the 
radio and X-ray light curves as synchrotron and IC emission from the secondary electrons produced in the
proton-proton interactions \citep{2007Ap&SS.309..253N}. 
However, observations show that the TeV flux from the system begins to increase $\sim50$~d before periastron, before the pulsar reaches the location of the disk suggested by the radio eclipse, which makes the hadronic scenario rather unlikely. 
In the leptonic scenario one would expect the light curve either to peak close to periastron in the case of dominate
adiabatic losses (due to the increased density of stellar photons and the most favorable scattering angle)
or to show a smooth variability in the case of the saturation of the electron spectrum by radiation losses \citep{1999APh....10...31K}. Neither of these predictions can explain the observed TeV light curve. In the case of the saturation regime a weak variability of the TeV flux is expected due to the orbital dependence
of the scattering angle, which will result in the flux being slightly higher before periastron (compared to the flux after periastron) with a peak
right before periastron where the scattering angle is the largest. 
\citet{2011A&A...525A..80K} suggested that the introduction of time-dependent adiabatic losses, which are dominant over the whole orbit, could explain the observed TeV light curve.
However, that study focused more on determining the profile of the time-dependent adiabatic cooling coefficient from the observational data, rather
than explaining the physical reasons for such a profile. 

The influence of the circumstellar disk on the non-thermal emission has also been investigated in smooth particle hydrodynamic (SPH) simulations undertaken by \citet{2012ApJ...750...70T} which were able to roughly reproduce the X-ray light curve using a very dense disk ($\sim10^{-9}$g\,cm$^{-3}$; denser than typical). Here the pre- and post-periastron peaks in the X-ray and TeV light curves were interpreted as the pulsar spin-down power being converted to particle acceleration more efficiently
when the pulsar passes through the disk.  However, the model did not reproduce the GeV and TeV light curves (assuming IC emission) and predicted a peak in the flux around periastron which is in contradiction with the observations.

The decrease in the TeV flux at periastron can be naturally explained by \gaga\ absorption caused by stellar and disk photons. 
Indeed, the geometry of the system infers that \gaga\ absorption would be most effective just before periastron since at that point the path of the emitted \gammaray\ photon towards the observer, will pass at the closest distance to the star (and therefore through a higher density of photons) and at the most optimal interaction angle.
Gamma-gamma absorption in \gammaray\ binaries, and in particular in \psrb, was thoroughly studied by
\citet{2006A&A...451....9D} concluding that although \gaga\ absorption should provide a significant effect on the observed TeV light curve from
\psrb, this effect alone is not sufficient to explain the shape of the light curve. However, this study
did not take into account the circumstellar disk of the Be-star which provides an additional photon field for \gaga\ absorption.
In addition to this, improved stellar parameters for \cstar\ have been obtained by \citet{2011ApJ...732L..11N}, which shows that the star has a higher effective temperature than previously thought. This in turn implies a higher photon energy density and hence stronger \gaga\ absorption.
Gamma-gamma absorption in the disk was briefly discussed in \citet{2014JHEAp...3...18S}, where it was shown
that the disk might significantly contribute to the overall \gaga\ absorption. However, \citet{2014JHEAp...3...18S} used a simplified
 description of the disk and assumed it had a constant width and a constant energy density. 
The maximum possible value for the energy density of the disk, for which the \fermi\ upper limits  were not violated by the intrinsic flux and cascade emission, was used.
 For this value, the  \gaga\ absorption is stronger in the disc than in the stellar radiation field. Therefore, the calculated flux
 extinction in \citet{2014JHEAp...3...18S} should only be considered as an upper limit for the stellar and binary parameters used in that work.

In this paper we revisit the problem of \gaga\ absorption in \psrb\ using a more realistic model of the circumstellar disk, with the energy density constrained by 
  infrared and optical observations of the star, and adopt the more updated stellar and binary parameters.

\section{Astrophysical properties of \psrb} \label{sec:model_setup}

\subsection{Astrophysical parameters} \label{sec:params}

Long-term observations of the radio pulsar provide very accurate measurements of \psrb's orbital parameters. 
The pulsar (\pulsar) is orbiting the  Be star \cstar\ in a very eccentric orbit ($e = 0.87$) with an orbital period of $1236.72$\,d, with a longitude of periastron of $\omega = 138.7^\circ$ \citep{1992MNRAS.255..401J, 1992ApJ...387L..37J,1994MNRAS.268..430J,2014MNRAS.437.3255S}. The binary system is at a distance of $\approx2.3$\,kpc from the Earth \citep{2011ApJ...732L..11N}. The last periastron took place on 2014 May 4 (MJD 56781.418307).

High-resolution optical spectroscopy of \cstar\ shows that the star rotates faster and is more luminous than it was previously thought, is oblate ($R_{{\mathrm{eq}}} \simeq 9.7 R_\odot$ at the equator and $R_{\mathrm{pole}} \simeq 8.1 R_\odot$ at the poles), with a temperature gradient from $T_{\mathrm{eq}} \simeq 27\,500$\,K to $T_{\mathrm{pole}} \simeq 34\,000$\,K.  \citep{2011ApJ...732L..11N}. 
For a non-rotating star, the equivalent stellar parameters are $R_\ast = 9.2 R_\odot$ and an effective temperature of $T_\ast = 33\,500$\,K \citep{2011ApJ...732L..11N}.
The mass of the star was estimated to be $M_\ast = 31 M_\odot$, however, it is noted by the authors that the uncertainty in this could be large since it is dependent on the estimated distance to the source as well as the corrections that must be applied due to stellar rotation \citep{2011ApJ...732L..11N}.  

\begin{figure}
\centering
\resizebox{\hsize}{!}{\includegraphics{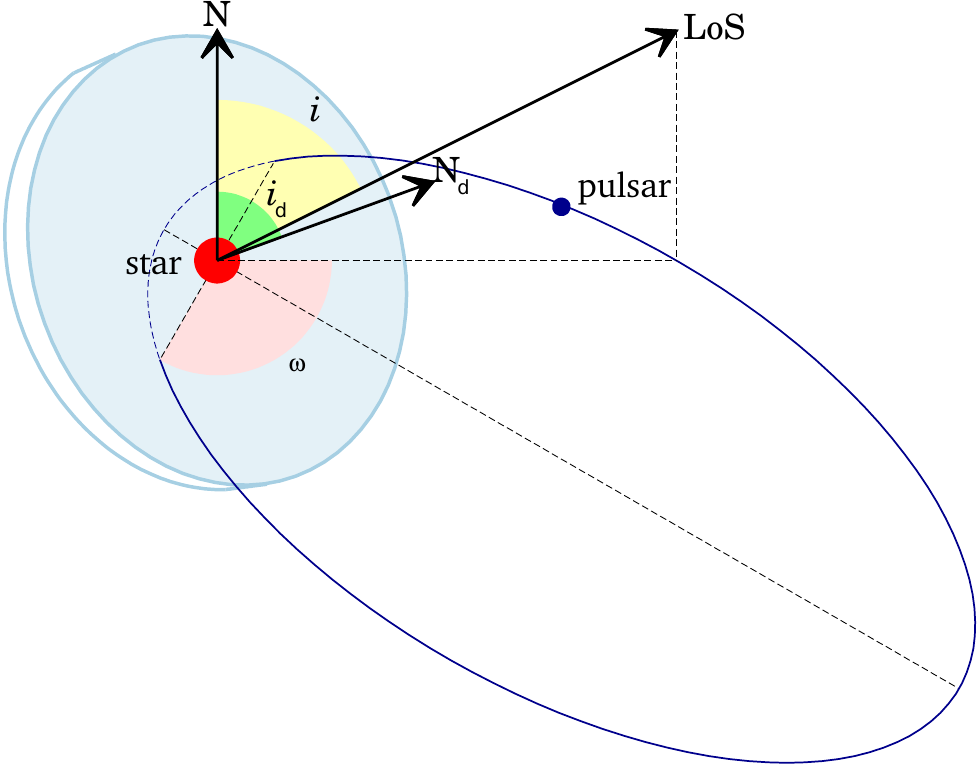}}
\caption{Schematic plot describing geometry of the binary system \psrb. N represents the normal to the orbital plane and N$_{\rm d}$ is the normal to the plane of the disk. The inclination of the binary $i$, the inclination of the disk $i_{\rm d}$, and the longitude of periastron $\omega$ are also shown. The pulsar is moving clockwise in this plot. }
\label{geometry}
\end{figure}

Figure \ref{geometry} illustrates a schematic picture of the binary system denoting the parameters which determine the complex geometry
of the system and its spatial orientation. Assuming the mass of the star is $M_\ast = 31 M_\odot$ (with a radius $R_\ast=9.2R_\odot$), the mass of the pulsar is $1.4 M_\odot$, and that the binary mass function is $f = 1.53 M_\odot$ \citep{1994MNRAS.268..430J},  the orbital inclination angle of the system is $i = 22.2^\circ$ and at periastron, the binary separation is $\simeq 21.9R_\ast$.

The inclination of the disk with respect to the orbit is argued to be small; approximately $i_{\mathrm d} = 10^{\circ}$ or less \citep{1995MNRAS.275..381M}.
The lower limit on the radius of the disk is $\sim45R_\ast$ determined by the separation distance at $\sim20$ days from periastron. Hereafter the radius of the disk is assumed to be $R_{\mathrm{disk}} = 50 R_\ast$. Circumstellar
disks of Be stars are believed to be thin \citep[see e.g.][]{2013A&ARv..21...69R} and following other theoretical works on this system
\citep[e.g.][]{2011PASJ...63..893O, 2012ApJ...750...70T, 2012MNRAS.426.3135V, 2014JHEAp...3...18S} we assume that the half-opening angle of the disk is
$\theta_{\mathrm disk} = 1^\circ$ throughout the paper.

\subsection{Radiation field} \label{sec:target_photons}

The Be star \cstar\ produces the radiation field which provides the target photons for inverse Compton scattering as well as for \gaga\ absorption of TeV \gammaray\ photons.
The radiation field consists of two components, the stellar component from the star as well as a secondary component from the circumstellar disk.

The circumstellar disks of Be stars are known to produce infrared radiation through free-free and free-bound scattering. We model the contribution from the disk following \citet{1986A&A...162..121W} where the disk is assumed to have a density profile given by,
\begin{equation}
 \rho(r) = \rho_0 \left( \frac{r}{R_\ast} \right)^{-n},
\end{equation}
where $r$ is the distance from the center of the star, $R_\ast$ is the radius of the star and $\rho_0$ is the density at the base of the disk (at $r=R_\ast$). The density decreases with the profile index $n$. In this model it is assumed that the disk has a uniform temperature, $T_{\rm disk}$, a half-opening angle, $\theta_{\rm disk}$ and has a maximum radius, $R_{\rm disk}$. 

The emission from the disk is calculated from the optical depth due to free-free and free-bound scattering. It can be shown that the optical depth along a length, $s$, through the disk is given by \citep{1984A&A...136...37L, 1986A&A...162..121W,2012MNRAS.426.3135V}
\begin{equation}
\tau_\nu =  \int_0^{\bar{s}} {\rm d}\bar{s}\,  X_\lambda X_\ast \bar{r}^{-2n} \, ,
\label{eqn:tau_nu_bar}
\end{equation}
where the barred distances are in units of stellar radii.
Here, 
% \begin{equation}
%  X_\lambda= \lambda^2 \left[ ( 1 - e^{-h\nu/kT_{\rm disk}} )/(h\nu/kT_{\rm disk})\right] \left[ g(\nu,T_{\rm disk}) + b(\nu,T_{\rm disk})\right],
% \label{eqn:x_lambda}
% \end{equation}
\begin{eqnarray}
\nonumber X_\lambda= \lambda^2 \left[ ( 1 - e^{-h\nu/kT_{\rm disk}} )/(h\nu/kT_{\rm disk})\right] \times \\ \left[ g(\nu,T_{\rm disk}) + b(\nu,T_{\rm disk})\right],
\label{eqn:x_lambda}
\end{eqnarray}
contains the wavelength dependent terms which vary with wavelength $\lambda$ (and frequency $\nu$), $k$ is the Boltzmann constant, and  $g(\nu,T_{\rm disk})$ and $b(\nu,T_{\rm disk})$ are the gaunt factors for the free-free and free-bound scattering, respectively. The gaunt factors are calculated following the approximation outlined in \citet{1984A&AS...57..327W} and for the analysis presented here we have only considered the contribution of the free-free scattering.
The term $X_\ast$ is given by
\begin{equation}
 X_\ast = 4.923 \times 10^{35} \, \overline{z^2} T_{\rm disk}^{-3/2} \mu^{-2}  \varpi \rho_0^2 \left(\frac{R_*}{R_\sun}\right),
\end{equation}
where $\overline{z^2}$ is the mean of the squared atomic charge, $\mu$ is the mean atomic weight and $\varpi$ is the ratio of the number of electrons to the number of ions.

The parameters for the disk model are taken from the fit to the stellar and disk component used in \citet{2012MNRAS.426.3135V} which were fitted to near-infrared and mid-infrared observations of LS~2883. The parameters are summarized in Table~\ref{tab:cog_parameters}.

 \floattable
\begin{deluxetable}{LC}
\tablecaption{Parameters for the disk model. \label{tab:cog_parameters}}
\tablecolumns{2}
% \tablenum{2}
 \tablewidth{20pt}
\tablehead{
\colhead{Parameter} &
\colhead{Value}}
\startdata
T_\ast & 33\,000~{\rm K} \\
n & 3.055 \\
\log_{10} X_\ast & 10.245 \\
R_{\rm disk} & 50 R_\ast \\
T_{\rm disk} & 19\,800~{\rm K} \\
\theta_{\rm disk} & 1\degr \\
\enddata
\end{deluxetable}
%\vspace{5mm}

Since, in this model, the disk is assumed to be in local thermodynamic equilibrium, the source function is given by the Planck function. The photon number density is therefore given by
\begin{eqnarray}
 n_{\rm disk}(\nu, \Omega) &=& \frac{I_\nu(\nu, \Omega)}{h \nu c} \\
&=& \frac{1}{h \nu c} B_\nu(T_{\rm disk}) \left[1-{\rm e}^{-\tau_\nu}\right],
\label{eqn:n_disc_eps}
\end{eqnarray}
where $h$ is the Planck constant, $c$ is the speed of light, and $B_\nu(T_{\rm disk})$ is the Planck function. 
 
The photon number density is calculated from equation~(\ref{eqn:n_disc_eps}) in any direction by determining the optical depth in that direction due to the circumstellar disk. The optical depth is calculated by numerically integrating equation~(\ref{eqn:tau_nu_bar}) over $\bar{s}$, taking into account the geometry of the circumstellar disk. The code also determines if any direction would intercept the star and correctly takes into account which regions of the disk are obscured and must not be included.

\begin{figure}
\centering
\resizebox{\hsize}{!}{\includegraphics{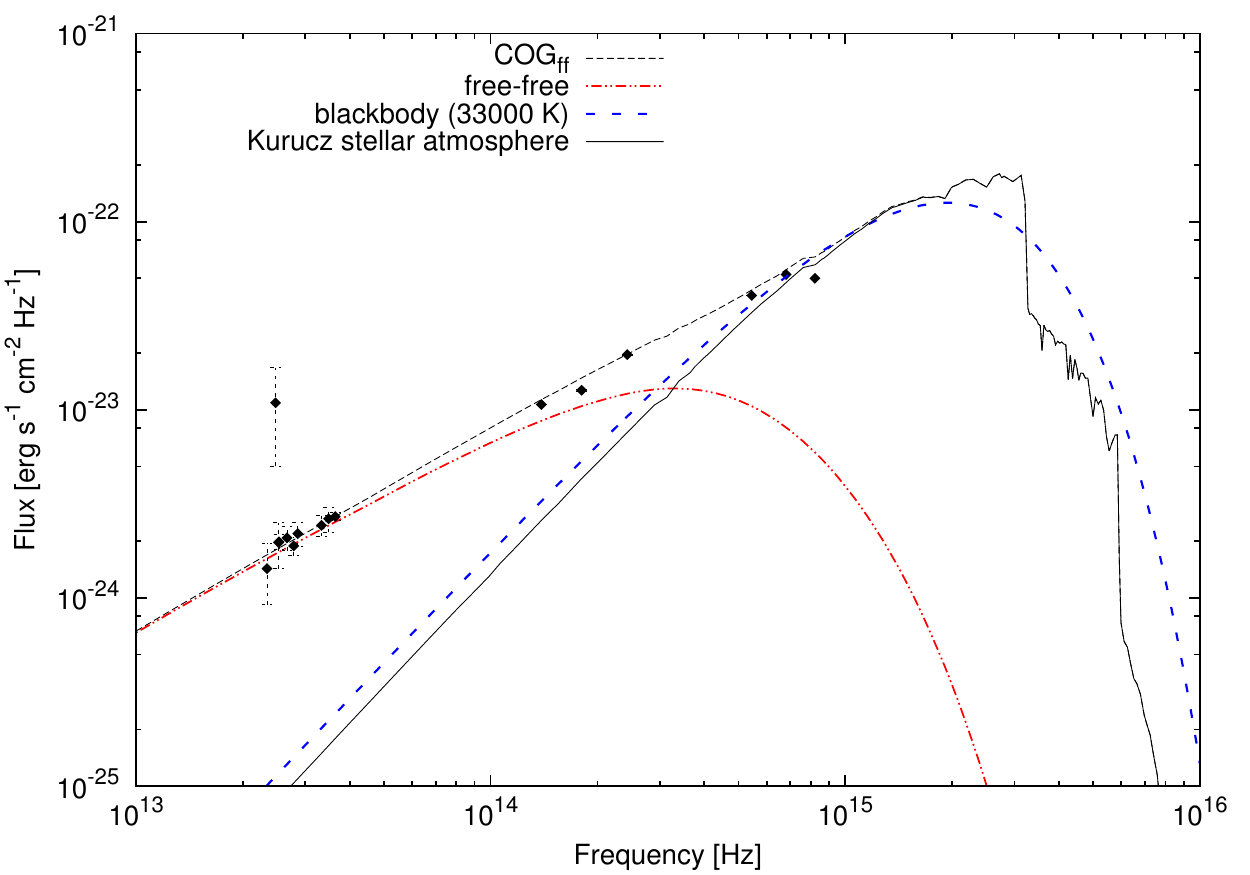}}
\caption{The infrared and optical observations from \cstar\ fitted with a stellar atmosphere (solid black line) and free-emission (red double-dot dashed line), along with the combined contribution (black dashed) as given in \citet{2012MNRAS.426.3135V}. A 33\,000\,K blackbody distribution is plotted for comparison (blue dashed line).   }
\label{fig:cog} 
\end{figure}

The contribution from the star has been modeled assuming the star is a spherical blackbody emitter, with a temperature of 33\,000\,K and a radius of $R_\ast = 9.2$~R$_\sun$ in order to be comparable to the previous model \citep{2012MNRAS.426.3135V}.  Since the star is surrounded by the circumstellar disk, the disk will also attenuate emission from the star. Therefore, similarly to above, the number density from the star is calculated as
\begin{equation}
 n_\ast (\nu,\Omega) = \frac{1}{h \nu c} B_\nu(T_\ast) {\rm e}^{-\tau_\nu},
\end{equation}
where the stellar contribution is decreased by $\exp(-\tau_\nu)$ if it is observed through the disk.
The model, compared to the observations, is shown in Fig.~\ref{fig:cog}.

\section{Gamma-gamma absorption} \label{sec:gg_absorption}

Gamma-ray photons emitted in binary systems are subject to $\gamma\gamma$ absorption as they pass through the star's photon field \citep[see e.g.][]{2006A&A...451....9D}.
The interaction of a \gammaray\ photon with a low-energy photon can result in electron-positron pair production \citep{1967PhRv..155.1404G}, if the energy exceeds the threshold condition for pair production which is given by
\begin{equation}
  \epsilon \epsilon_\gamma (1-\cos \theta) \leq 2,
  \label{eqn:gaga_threshold}
\end{equation}
where $\epsilon$ and $\epsilon_\gamma$ denote the photon energies of the target low-energy photon and the \gammaray\ photon, respectively, normalized to the electron rest-mass energy, i.e. $\epsilon = h\nu/m_{\rm e}c^2$, and $\theta$ is the interaction angle between the two photons. The minimum threshold occurs for a head-on collision ($\cos \theta = -1$), and for a 1\,TeV gamma-ray photon this will require photons with a threshold frequency of,
\begin{equation}
 \nu \simeq 6.3\times10^{13} \left(\frac{h \nu_\gamma }{1~{\rm TeV}}\right)^{-1}~{\rm Hz}
\end{equation}
which is within the mid-infrared regime (where the disk is the primary photon source, Fig.~\ref{fig:cog})

The \gaga\ optical depth is given by \citep{1967PhRv..155.1404G}
\begin{equation}
  \tau_{\gamma\gamma} = \int_0^l\,{\rm d}l \int_{4\pi}  {\rm d}\Omega \, \,(1-\mu) \int_{\frac{2}{\epsilon_\gamma (1-\mu)}}^\infty {\rm d} \epsilon \, n_{\rm ph}(\epsilon,\Omega) \sigma_{\gamma\gamma}(\epsilon,\epsilon_\gamma,\mu)
\label{eqn:tau_gamma}
\end{equation}
where $l$ is the distance over which the \gammaray\ photon travels, 
$\mu = \cos \theta$, ${\rm d}\Omega = {\rm d}\mu {\rm d}\phi$, and $n_{\rm ph}(\epsilon,\Omega)$ is the number density of the low-energy target photons. Here, $\sigma_{\gamma\gamma}$, is the \gaga\ cross-section \citep{1976tper.book.....J},
\begin{equation}
 \sigma_{\gamma\gamma}(\beta) = \frac{3}{16} \sigma_{\rm T} ( 1- \beta^2) \left[ (3-\beta^4) \ln \left(\frac{1+\beta}{1-\beta} \right) -2\beta (2-\beta^2)\right],
\end{equation}
where 
\begin{equation}
 \beta = \sqrt{1-\frac{2}{\epsilon \epsilon_\gamma (1-\mu)}},
\end{equation}
and $\sigma_{\rm T}$ is the Thomson cross-section. 
The maximum \gaga\ cross-section occurs when the energy of the interacting photons is twice that of the energy threshold (equation~\ref{eqn:gaga_threshold}). This implies that for TeV \gammaray\ photons the \gaga\ interaction is a maximum for infrared photons at a frequency  
\begin{equation}
   \nu \simeq 1.3\times10^{14} \left(\frac{h \nu_\gamma }{1~{\rm TeV}}\right)^{-1}~{\rm Hz}.
\end{equation}

\begin{figure}
\centering
\resizebox{\hsize}{!}{\includegraphics{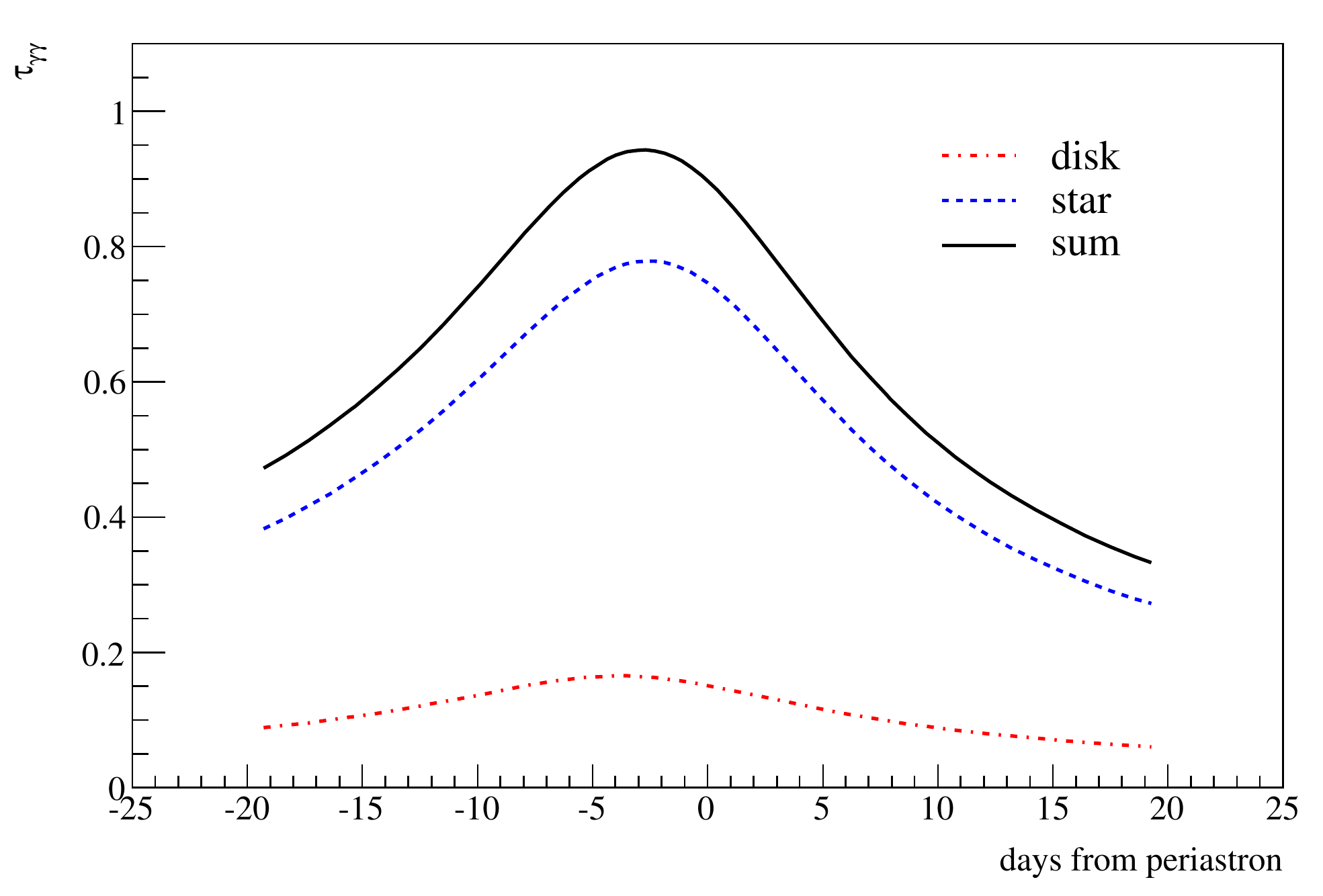}}
\caption{The $\gamma\gamma$ optical depth of the Be star's radiation field as a function of time around periastron for a 1\,TeV \gammaray\ photon. The blue dashed line denotes the stellar component, the red dot-dashed line denotes the circumstellar disk component and the combined effect is shown by the solid black line.}
\label{opacity}
\end{figure}

We treat the two components of the target radiation field (stellar radiation and circumstellar disk radiation) separately, and numerically calculate
the \gaga\ optical depth of both components for a \gammaray\ photon traveling in the direction of the observer. The total optical depth is given by the sum of the optical depths of the two components. Fig. \ref{opacity} shows the optical depth
for a \gammaray\ photon with an energy of 1\,TeV as a function of time from periastron.  As expected the \gaga\ optical depth due to the photons from the disk
is lower than that due to the star but still results in a $\approx15\%$ decrease of the flux at TeV energies. The total decrease of the flux at 1\,TeV due to \gaga\,absorption may reach $\sim60\%$. For the geometrical configuration used in our calculations
(see Section \ref{sec:params}) the maximum absorption due to the disk is at 
$\approx4$\,d prior to periastron and
due to the star at $\approx2$\,d before periastron.  

\section{TeV emission}

We assume that the TeV emission from \psrb\ is produced by relativistic particles accelerated at the termination shock that occurs between the pulsar and stellar winds. The location of the termination shock depends on the ratio of the pulsar to stellar wind power
and on the binary separation distance.  
Moreover, the mass-loss rate and velocity of the stellar wind are different in the
equatorial and polar regions. Therefore, the location and morphology of the shock changes across the orbit. However, close
to periastron, given that the inclination angle of the disk is rather small ($i_{\rm d}\sim10^\circ$), the environment around the pulsar is strongly influenced by the dense equatorial disk and the termination shock should form close to the pulsar \citep[see discussion in][]{2014JHEAp...3...18S}. 
Therefore, we assume the source of the TeV emission to be point-like, and neglect the distance between the pulsar and the termination shock. In Section \ref{sec:discussion}, we will briefly discuss how our results depend on the location of the termination shock.

\begin{figure}
\centering
\resizebox{\hsize}{!}{\includegraphics{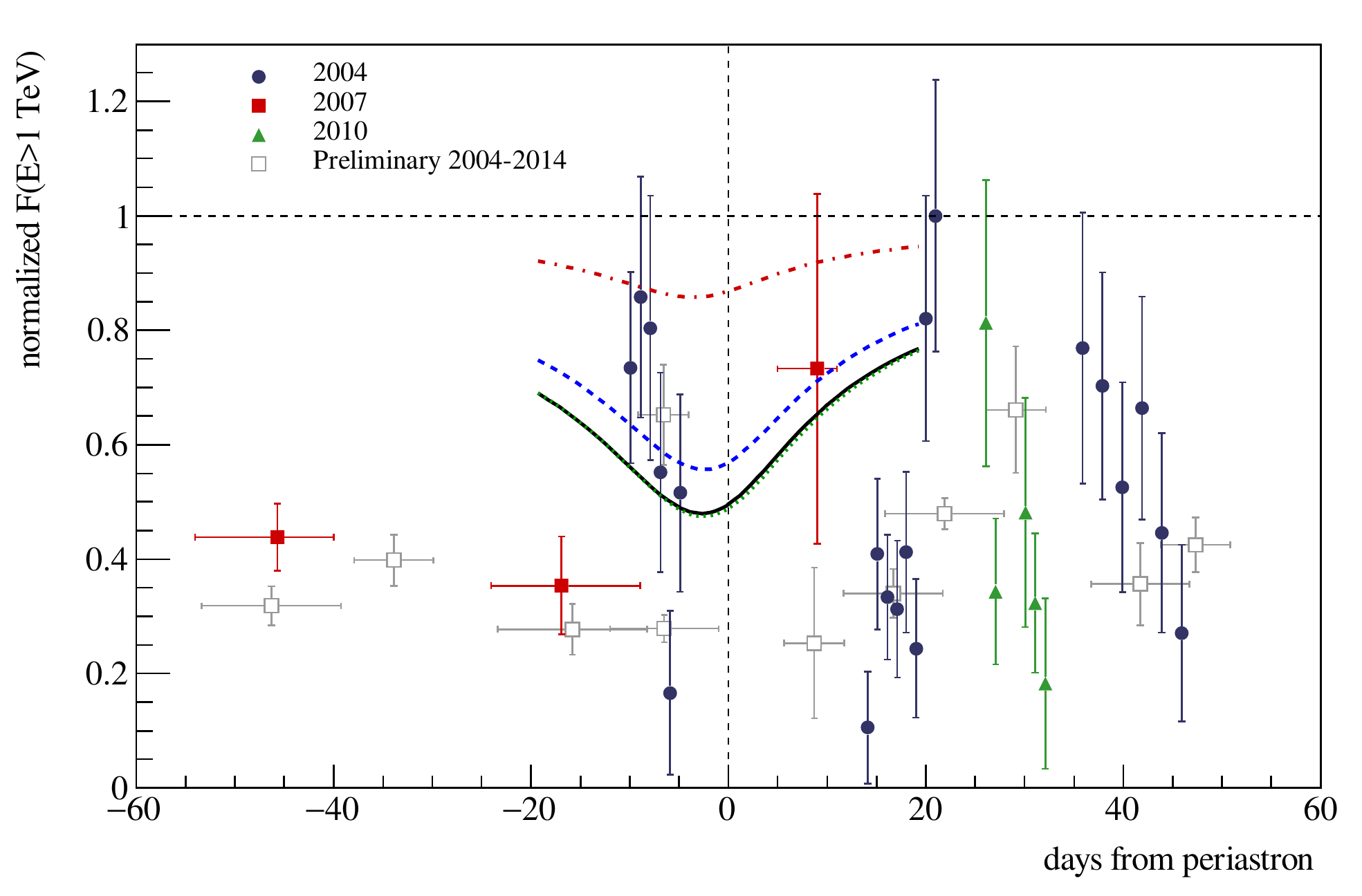}}
\caption{The flux ($>1$\,TeV) around periastron assuming the underlying intrinsic flux is constant, with a photon index $\Gamma=3.0$. The plot shows the absorption due to disk photons (red dot-dashed line), the stellar photons (blue dashed line) and the combined effect (solid black line). The H.E.S.S. data (colored, closed points) are from \citet{2013A&A...551A..94H}, while the open squares show the preliminary data reported by \citet{2015arXiv150903090R}. The dotted green line shows the \gaga\ absorption due to the stellar photons if the emission is produced at the apex of bow shock as discussed in Section~\ref{sec:discussion}. }
\label{fig:hess_plot} 
\end{figure}

In order to determine how the TeV observations will be affected by the \gaga\ absorption we have modeled the TeV light curve around periastron assuming an intrinsic power-law photon distribution,
\begin{equation}
\frac{{\rm d}N}{{\rm d}E}  = N_0 \left( \frac{E}{1\,{\rm TeV}}\right)^{-\Gamma} 
\end{equation}
with a photon index of $\Gamma=3.0$. For the region around periastron we assume a constant flux, and that the variation is only due to the change in \gaga\ absorption. In Fig.~\ref{fig:hess_plot} the results are compared to the gamma-ray flux detected by H.E.S.S.\ ($>1$\,TeV) normalized to the highest flux value.

The \gaga\ absorption from the disk photons alone result in a decrease of $\approx14$\% which is a maximum at $\approx4$\,d before periastron. Similarly, including the stellar contribution results in a maximum  total absorption of $\approx52$\% at $\approx2$\,d before periastron. This is comparable to the results calculated for a 1\,TeV photon.

\begin{figure}
\centering
\resizebox{\hsize}{!}{\includegraphics{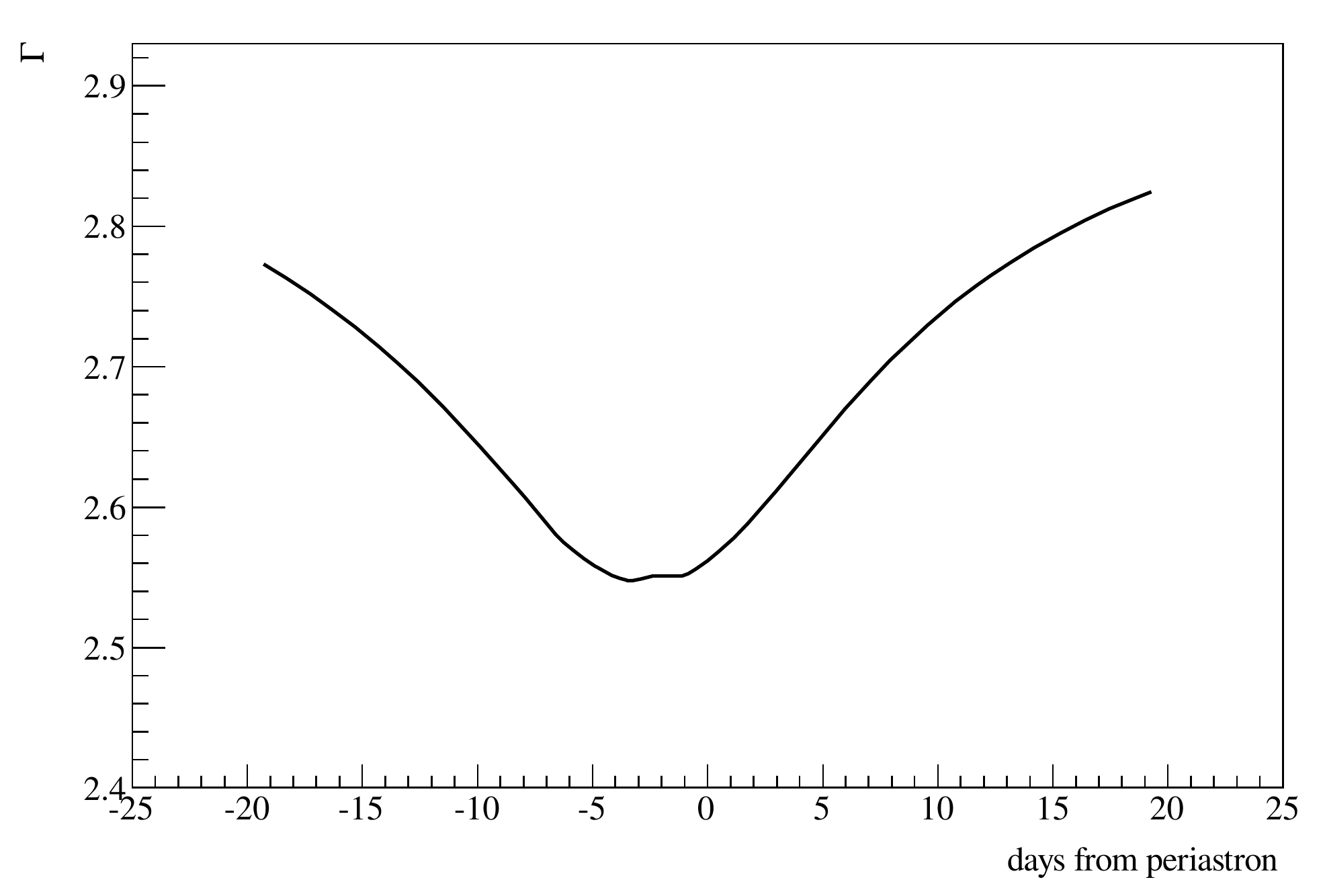}}
\caption{Variation in the gamma-ray photon index in the energy range $1-50$\,TeV around periastron due to \gaga\ absorption assuming an intrinsic photon index of $\Gamma=3.0$.} 
\label{fig:photon_index} 
\end{figure}

The energy dependence of the \gaga\ absorption also results in a variation of the photon index around periastron. We determine how the TeV $\gamma$-ray photon index changes by calculating the \gaga\ absorption for an intrinsic power-law photon distribution ($\Gamma=3.0$) and find the best-fit power-law distribution for the absorbed spectrum in the 1\,TeV to 50\,TeV energy range. Fig.~\ref{fig:photon_index} shows the photon index variability around periastron. The photon index varies between $\sim 2.6-2.8$ becoming harder near periastron. This result is comparable with the photon index ($\Gamma\approx 2.8$) measured by H.E.S.S.\  away from periastron.

\section{Discussion}
\label{sec:discussion}

\subsection{TeV light curve}
The results of the \gaga\ absorption, taking into account both the disk and stellar component (with the updated stellar parameters), show that it will have a significant effect on the observed TeV light curve. Gamma-gamma absorption causes a significant decrease in the flux within a few days from periastron which is  co-incident with a 
dip in the observed flux. Note that the observed flux shown in Fig.~\ref{fig:hess_plot} is normalized to the highest flux data point which reflects the daily averaged flux $\sim20$\,days after the 2004 periastron. Preliminary results of the re-analysis of all the H.E.S.S. data \citep{2015arXiv150903090R} show period-averaged fluxes with absolute values somewhat lower (but still compatible within errors) than the daily averaged fluxes from the 2004 periastron (Fig.~\ref{fig:hess_plot}). Compared to the re-analysed data only (gray open squares in Fig.~\ref{fig:hess_plot}), the modeled light curve  can reproduce not only the location of the dip but also its depth.
However, \gaga\ absorption alone is not sufficient to explain the full complexity of the observed TeV light curve around periastron. For example the steep decrease of the observed flux before periastron is not compatible with the rather smooth decrease predicted by \gaga\ absorption. Therefore, some variable features must be present already in the intrinsic flux from the source. In this study we do not model the emission from the source nor do we consider the complicated morphology of the termination shock (see below), which might further modify the TeV light curve. Instead we assume that the intrinsic TeV emission is constant close to periastron and  that the emitting region is point-like, in order to isolate the effect of the \gaga\ absorption and its impact on the resulting light curve.

\begin{figure}
\centering
\resizebox{\hsize}{!}{\includegraphics{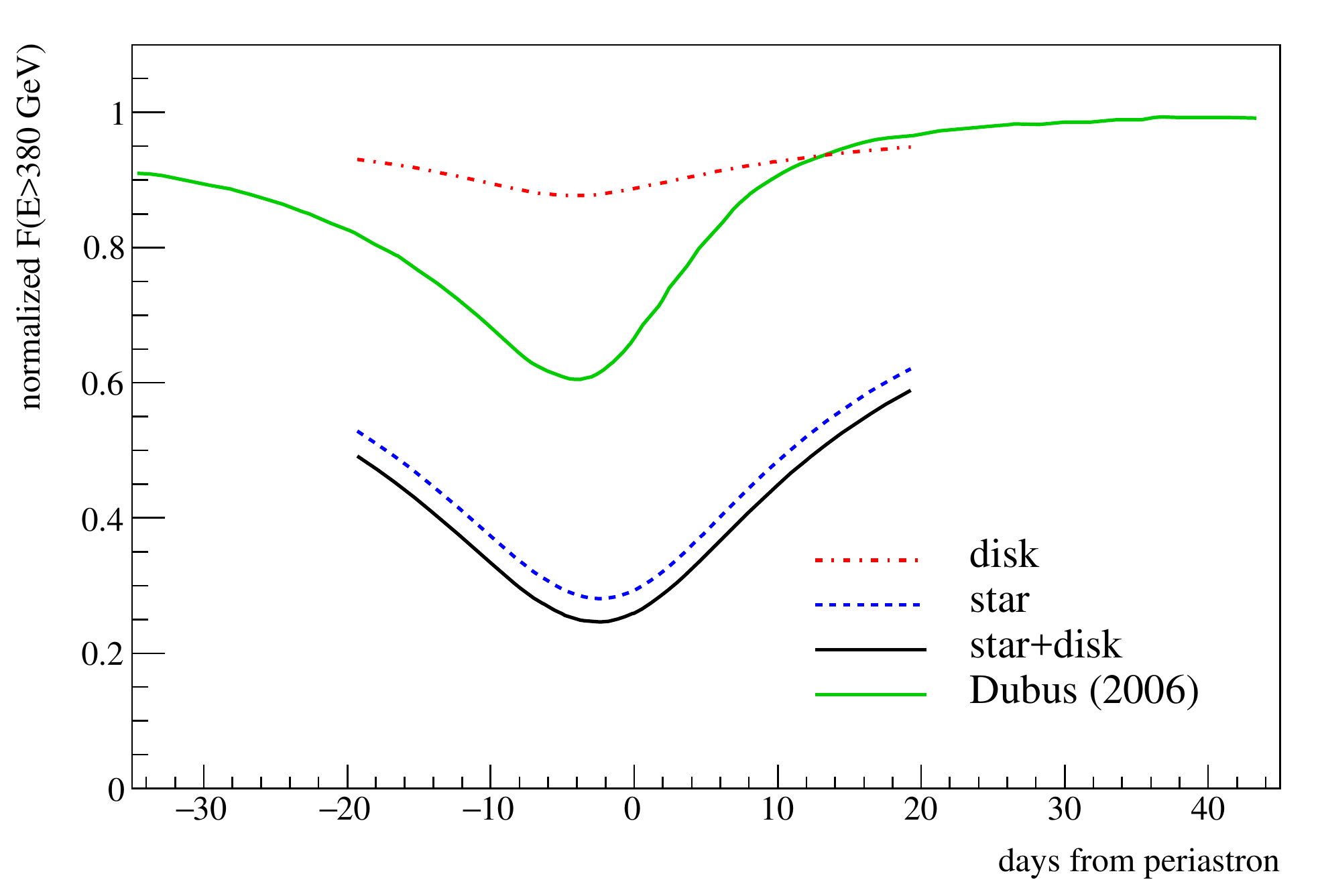}}
\caption{The integrated \gaga\ absorption above 380\,GeV due to the photons from the star and circumstellar disk assuming a power-law photon distribution with $\Gamma=2.8$. The green solid line shows the \gaga\ absorption found in \citet{2006A&A...451....9D} compared to the updated emission as found here for the disk (dot-dashed red line) and star (dashed blue line). The solid black line shows the combined contribution.}
\label{fig:dubus_comparison_plot} 
\end{figure}

The effect of the \gaga\ absorption is larger than was found using the previous stellar parameters \citep{2006A&A...451....9D}
and the minimum in the light curve is also shifted slightly due to the updated orbital parameters. The increase in the absorption due to the updated stellar parameters is
more apparent at lower energies. Fig.~\ref{fig:dubus_comparison_plot} shows the integrated \gaga\ absorption above 380\,GeV, assuming an intrinsic photon index of
$\Gamma=2.8$, in order to make a more direct comparison to the results by \citet{2006A&A...451....9D}. The combined star and disk contribution leads to a maximum
integrated \gaga\ absorption of $\sim75$\%.

The disk contribution to the overall \gaga\ absorption is much smaller than was suggested in \citet{2014JHEAp...3...18S}.
  This is, however, not surprising as there the \gaga\ absorption in the disk was calculated 
  for the highest possible disk energy density for which the \fermi\ upper limits were not violated 
  while in this work  the disk radiation field is constrained by infrared observations of the star. 
  The different shape of the orbital dependent \gaga\ absorption is due to the more realistic description of the
  disk presented in this work compared to the simplified assumptions of constant width and density
  applied in \citet{2014JHEAp...3...18S}.   

While our model has accurately taken into account the full geometric effect of the star and the circumstellar disk, for simplicity the emitting region has been considered as a point source located at the position of the pulsar. However, hydrodynamic simulations of colliding stellar and pulsar winds in binary systems show that the orbital evolution of the wind interaction results in a very complicated shock morphology with strongly entangled features \citep[see][and references therein]{2015A&A...577A..89B}. It is unclear where relativistic particles and, subsequently, \gammaray\ emission are produced. If the TeV emission is produced far from the apex of the shock, \gaga\ absorption might be negligible since the path of the emitted \gammaray\ photon will lie far from the star and the radiation density  will be much lower.
Contrary to this, if the TeV emission is produced near the apex of the shock, TeV photons will pass
  closer to the star and the higher optical energy density will result in a larger \gaga\ optical depth. 
  The distance from the pulsar to the apex of the shock is given by \citep{1993ApJ...402..271E}
  \begin{equation}
    r_{\rm p} = d\frac{\sqrt{\eta}}{(1+\sqrt{\eta})},
  \end{equation}
  where $d$ is the distance between the pulsar and the star and $\eta$ is the winds' ram pressure ratio, which is estimated to be $\eta = (0.03 - 0.08)$ for the polar wind
  \citep[see][and references therein]{2014JHEAp...3...18S}. In this case
  the maximum distance from the pulsar to the apex of the shock %in the apex 
  is $r_{\rm p}  = 0.22d$. The green dotted curve in Fig.\,\ref{fig:hess_plot} shows the absorption above 1\,TeV due only to the stellar (and not the disc) photons under the
  assumption that the TeV emission is produced at the apex of the shock and that the distance to the apex is a constant fraction of the separation distance, $r_{\rm p}  = 0.22d$. 
  This can be considered as an upper limit for the \gaga\ absorption
  in the system.
  The absorption is similar to that found by the more detailed analysis which also considered the \gaga\ absorption due to the disk, but assumed the emission originated at the position of the pulsar.  Future observations with the much more sensitive CTA {\citep{2011ExA....32..193A}} combined with the predicted extinction
  of flux due to \gaga\ absorption might provide a hint of the location of the TeV \gammaray\ emission in the system.

\subsection{Spectral energy distribution}

\begin{figure}
\centering
\resizebox{\hsize}{!}{\includegraphics{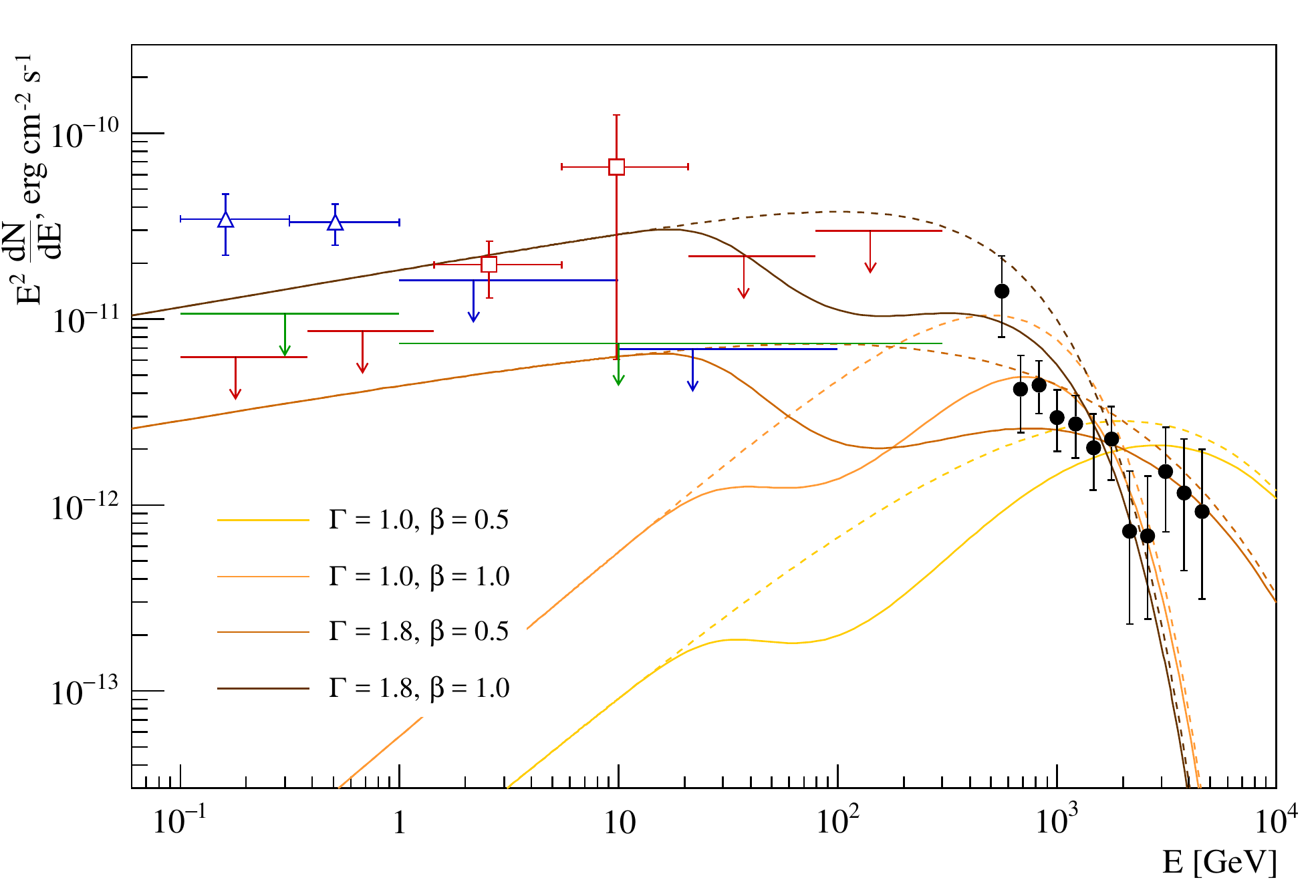}}
\caption{Spectral energy distribution of \psrb\ with the associated photon distributions discussed in the text.  The {\it Fermi}-LAT results from the 2010 periastron are taken from \citet[][blue triangles]{2011ApJ...736L..11A} and from \citet[][red squares]{2011ApJ...736L..10T}, and from the 2014 periastron from \citet[][green upper-limits]{2015ApJ...798L..26T}. The H.E.S.S.\ data are from the 2010 periastron \citep{2013A&A...551A..94H}. The intrinsic (dashed lines) and \gaga\ absorbed (solid lines) photon spectra are plotted for different parameters.}
\label{fig:SED} 
\end{figure}

In order to investigate how \gaga\ absorption will affect the observed $\gamma$-ray spectral energy distribution,
we have calculated the absorbed $\sim0.1-10^4$\,GeV spectrum assuming the underlying photon spectrum is given by a power-law with a super-exponential cutoff,
\begin{equation} 
 \frac{{\rm d}N}{{\rm d}E} = N_0 \left( \frac{E}{1\,{\rm TeV}} \right)^{-\Gamma} \exp \left[ -\left(\frac{E}{E_{\rm c}}\right)^\beta \right],
\end{equation}
which is a typical distribution for $\gamma$ rays produced via inverse Compton scattering.
Since the numerical calculation of the contribution from the disk is very computationally expensive, and the \gaga\ absorption is dominated by the stellar photons (particularly below 1\,TeV), we have calculated the \gaga\ absorption at periastron,  only considering the stellar contribution. The model is compared to the TeV observations $\sim30$\,d after periastron, which is comparable to the average flux detected around periastron, and is normalized assuming a flux of $F(>1\,{\rm TeV}) = 1.01\times10^{-12}$\,cm$^{-2}$\,s$^{-1}$ \citep{2013A&A...551A..94H}. 
The photon spectrum for $\Gamma = 1.0$ \& $1.8$, and for $\beta=0.5$ \& $1.0$, with a constant cutoff energy of $E_{\rm c}=500$\,GeV, are shown in Fig.~\ref{fig:SED}.\footnote{Please note that this choice of $\Gamma$ is chosen to correspond to the spectrum of the broad \gammaray\ range and is different to the photon spectrum only in the {\it Fermi} or H.E.S.S.\ energy range.
This range of values for the photon index $\Gamma$ was considered to accommodate $\Gamma = 1.5$ which corresponds to the canonical value of the shock accelerated particle spectrum index of $2.0$.}
The \gaga\ absorption results in a  decrease around 100\,GeV which makes it easier to reconcile the H.E.S.S.\ observations with the {\it Fermi}-LAT 
observations close to periastron. While pair-cascading may slightly increase the emission in the GeV range, the effect will be too low to change these results \citep{2014JHEAp...3...18S}. 

\subsection{Variation in photon index }

As discussed above, the variation in the photon index (Fig.~\ref{fig:photon_index}) is compatible with observations away from periastron. However, the sensitivity of the current H.E.S.S.\ array is too low to detect the predicted variation at periastron. If this model is correct, there will be a hardening of the spectrum as the pulsar reaches periastron. Since the source is currently detectable with H.E.S.S., the improved sensitivity of CTA  \citep{2011ExA....32..193A} will allow for a better detection of the system around periastron and may allow for this change in photon index to be measurable. While observations are not possible during the following periastron (2017 September) they will be possible during the next, in 2021 February, when conditions will be far more favorable for attempting to observe this effect.

\section{Summary}

The binary system \psrb\ is part of the growing class of $\gamma$-ray binaries. The production of TeV $\gamma$-rays in this system is most commonly assumed to be produced through the acceleration of electrons in the shock that forms between the pulsar and stellar winds, and the subsequent cooling of these electrons through inverse Compton scattering. These high energy photons, exceed the threshold for pair-production which should lead to a decrease in the TeV regime through $\gamma\gamma$ absorption. 

Gamma-gamma absorption was previously investigated for the known systems by \citet{2006A&A...451....9D}, however, since then further observations have refined the binary and stellar parameters. We have re-investigated \gaga\ absorption in \psrb\ incorporating the newer stellar and binary parameters \citep[e.g.][]{2011ApJ...732L..11N,2014MNRAS.437.3255S}, as well as taking into account the contribution from the circumstellar disk. 

We have shown that the combined contribution of the circumstellar disk and star result in an extinction of $\sim52$\% of the very high energy $\gamma$-rays ($>1$\,TeV) a few days before periastron. This maximum in the absorption is consistent with the dip in the flux observed by the H.E.S.S.\ telescope array and, therefore, we suggest that \gaga\ absorption is a major contributing factor to the TeV light curve and also contributes to the non-detection of the source in the $\sim100$\,GeV energy range. The \gaga\ absorption will produce a hardening of the TeV spectrum around periastron, an effect that may be observable with CTA around the 2021 periastron passage. 

Lastly, the exact location of the production of the $\gamma$ rays in $\gamma$-ray binaries is still unclear. We have in addition calculated an upper-limit to the \gaga\ absorption (due the stellar photons alone) by also calculating the absorption if the $\gamma$ rays are produced at the apex of the bow shock assuming $r_{\rm p} =0.22d$. 
    This results in a maximum extinction of $\sim50$\% of  the flux above 1\,TeV. The additional contribution of the disk to the total absorption would be $\gtrsim14$\% suggesting that the maximum absorption could be $\gtrsim65$\% if the circumstellar disk is included. More sensitive observations with CTA may provide constraints on the magnitude of the \gaga\ absorption in this system, which, combined with these estimates, might hint at the location of the TeV \gammaray\ emission in this system as well as the other $\gamma$-ray binaries.

\acknowledgments
The authors thank Maxim Barkov for fruitful discussions. 
This work was supported by the Department of Science and Technology
and the National Research Foundation of South Africa through a block grant
to the South African Gamma-Ray Astronomy Consortium. The numerical calculation were performed using the University of the Free State High Performance Computing Unit. 
The authors thank Maxim Barkov for fruitful discussions.

\bibliographystyle{aasjournal} 
\bibliography{references_full}

%% \begin{thebibliography}{}

%% \bibitem[Corrales(2015)]{2015ApJ...805...23C} Corrales, L.\ 2015, \apj, 805, 23
%% \bibitem[Hanisch \& Biemesderfer(1989)]{1989BAAS...21..780H} Hanisch, R.~J., \& Biemesderfer, C.~D.\ 1989, \baas, 21, 780 
%% \bibitem[Lamport(1994)]{lamport94} Lamport, L. 1994, LaTeX: A Document Preparation System, 2nd Edition (Boston, Addison-Wesley Professional)
%% \bibitem[Schwarz et al.(2011)]{2011ApJS..197...31S} Schwarz, G.~J., Ness, J.-U., Osborne, J.~P., et al.\ 2011, \apjs, 197, 31  
%% \bibitem[Vogt et al.(2014)]{2014ApJ...793..127V} Vogt, F.~P.~A., Dopita, M.~A., Kewley, L.~J., et al.\ 2014, \apj, 793, 127  

%% \end{thebibliography}



\end{document}